\PassOptionsToPackage{prologue,table,xcdraw}{xcolor}
\documentclass[sigconf]{acmart}
\usepackage{enumitem}

\usepackage{amssymb}
\usepackage[ruled]{algorithm2e}
\usepackage{url}
\usepackage{multirow}
\usepackage{subfigure}
\usepackage{bm}
\usepackage{verbatim}
\usepackage{ragged2e}
\usepackage{caption}
\usepackage{subcaption}
\usepackage{graphicx}
\usepackage[normalem]{ulem}
\useunder{\uline}{\ul}{}
\usepackage{amsmath}
\usepackage{longtable}
\usepackage{adjustbox}
\usepackage{microtype}
\usepackage{setspace}
\usepackage{hyperref}
\usepackage{float}
\usepackage{balance}
\newcommand{\ie}{\emph{i.e., }}

\newcommand{\aka}

\newcommand{\bym}[1]{\textcolor{black}{#1}}
\newcommand{\sigir}[1]{\textcolor{black}{#1}}
\newcommand{\new}[1]{\textcolor{black}{#1}}

\AtBeginDocument{%
  \providecommand\BibTeX{{%
    \normalfont B\kern-0.5em{\scshape i\kern-0.25em b}\kern-0.8em\TeX}}}

\copyrightyear{2025}
\acmYear{2025}
\setcopyright{acmlicensed}
\acmConference[SIGIR '25]{Proceedings of the 48th International ACM SIGIR Conference on Research and Development in Information Retrieval}{July 13--18, 2025}{Padua, Italy}
\acmBooktitle{Proceedings of the 48th International ACM SIGIR Conference on Research and Development in Information Retrieval (SIGIR '25), July 13--18, 2025, Padua, Italy}
\acmDOI{10.1145/3726302.3730105}
\acmISBN{979-8-4007-1592-1/2025/07}

\begin{document}

\title{Unconstrained Monotonic Calibration of Predictions in Deep Ranking Systems}


\author{Yimeng Bai}
\orcid{0009-0008-8874-9409}
\affiliation{
  \institution{University of Science and Technology of China}
  \city{Hefei}
  \country{China}
}
\email{baiyimeng@mail.ustc.edu.cn}
\authornote{Work done at Kuaishou.}

\author{Shunyu Zhang}
\orcid{0009-0000-1936-1162}
\affiliation{
  \institution{Kuaishou Technology}
  \city{Beijing}
  \country{China}
}
\email{zhangshunyu@kuaishou.com}

\author{Yang Zhang}
\orcid{0000-0002-7863-5183}
\affiliation{
  \institution{National University of Singapore}
  \city{Singapore}
  \country{Singapore}
}
\email{zyang1580@gmail.com}
\authornote{Corresponding author.}

\author{Hu Liu}
\orcid{0000-0003-2225-7387}
\affiliation{
  \institution{Kuaishou Technology}
  \city{Beijing}
  \country{China}
}
\email{hooglecrystal@126.com}

\author{Wentian Bao}
\orcid{0009-0001-2195-0553}
\affiliation{
  \institution{Columbia University}
  \city{Beijing}
  \country{China}
}
\email{wb2328@columbia.edu}

\author{Enyun Yu}
\orcid{0009-0009-0847-7464}
\affiliation{
  \institution{Northeasten University}
  \city{Beijing}
  \country{China}
}
\email{yuenyun@126.com}

\author{Fuli Feng$^{\dag}$}
\orcid{0000-0002-5828-9842}
\affiliation{
  \institution{University of Science and Technology of China}
  \city{Hefei}
  \country{China}
}
\email{fulifeng93@gmail.com}

\author{Wenwu Ou}
\orcid{0009-0004-2437-6835}
\affiliation{
  \institution{Unaffiliated}
  \city{Beijing}
  \country{China}
}
\email{ouwenwu@gmail.com}

\def\authors{Yimeng Bai, Shunyu Zhang, Yang Zhang, Hu Liu, Wentian Bao, Enyun Yu, Fuli Feng, Wenwu Ou}

\renewcommand{\shortauthors}{Yimeng Bai et al.}

\begin{abstract}

Ranking models primarily focus on modeling the relative order of predictions while often neglecting the significance of the accuracy of their absolute values. However, accurate absolute values are essential for certain downstream tasks, necessitating the calibration of the original predictions. To address this, existing calibration approaches typically employ predefined transformation functions with order-preserving properties to adjust the original predictions. Unfortunately, these functions often adhere to fixed forms, such as piece-wise linear functions, which exhibit limited expressiveness and flexibility, thereby constraining their effectiveness in complex calibration scenarios. To mitigate this issue, we propose implementing a calibrator using an Unconstrained Monotonic Neural Network (UMNN), which can learn arbitrary monotonic functions with great modeling power. This approach significantly relaxes the constraints on the calibrator, improving its flexibility and expressiveness while avoiding excessively distorting the original predictions by requiring monotonicity. Furthermore, to optimize this highly flexible network for calibration, we introduce a novel additional loss function termed Smooth Calibration Loss (SCLoss), which aims to fulfill a necessary condition for achieving the ideal calibration state. Extensive offline experiments confirm the effectiveness of our method in achieving superior calibration performance. Moreover, deployment in Kuaishou's large-scale online video ranking system demonstrates that the method's calibration improvements translate into enhanced business metrics. The source code is available at \url{https://github.com/baiyimeng/UMC}. 

\end{abstract}

\ccsdesc[500]{Information systems~Information retrieval}

\keywords{Calibrator Modeling; Unconstrained Monotonic Neural Network; Ranking System}


\maketitle

\section{Introduction}\label{section:intro}

Ranking systems are paramount in recommendation and search, serving as the core function to effectively filter out user-interested content from the vast ocean of information~\cite{SIM, QIN, D2Q, TWIN}. These systems typically learn to predict user-item (or query-document) matching scores for sorting candidates. While the primary focus is on ensuring the accuracy of relative orders for ranking, the accuracy of absolute values is also essential for certain downstream tasks~\cite{SBCR,ScaleCalib}. For example, in computational advertising, the absolute values of matching scores (\ie the predicted click-through rate) are critical for expected revenue estimation in advertisement bidding~\cite{Bidding, AdaCalib}. This necessitates \textit{calibration} of predicted scores, ensuring that they are consistent with the actual likelihood when converted into probabilistic predictions~\cite{JRC,RCR,calibration}.

Existing works employ fixed-form order-preserving transformations for calibration while maintaining ranking performance. Early methods, such as binning~\cite{HB, MBCT}, isotonic regression~\cite{IR, SIR}, and scaling~\cite{PlattScaling, TemperatureScaling, BetaCalib, GammaGauss}, typically use the statistical characteristics of original predictions to perform univariate transformations that ensure global order preservation. Unfortunately, these transformations have limited parameter space, making the methods inherently ineffective for industrial deep ranking models~\cite{LiRank}. Later works explore the utilization of neural networks for calibration, using them to adaptively learn parameters of transformation functions for samples with different features~\cite{NeuralCalib,AdaCalib,SBCR,DESC}. These methods provide more adaptability and achieve multivariate transformation, however, most underlying functions still adhere to a fixed form, such as piece-wise linear functions, which exhibit inadequate fitting performance~\cite{error}. These excessive constraints on the architecture result in insufficient expressiveness, limiting their efficacy in scenarios like multi-field calibration that involves nuanced patterns~\cite{ConfCalib}. \bym{Moreover, uncalibrated value errors in specific fields can amplify within feedback loops, ultimately undermining the integrity of the entire recommendation ecosystem~\cite{DecRS}.}

Given the drawbacks of existing methods, we aim to reduce constraints on the calibrator architecture. This will enhance its expressiveness and flexibility, allowing for more accurate calibration that aligns the prediction of each sample more closely with its actual likelihood. Specifically, we may need to weaken the strict order-preserving requirements, as ideal calibration should also correct potential ranking errors, and allow different transformations for different samples. However, a suitable trade-off in reducing the constraints is necessary; otherwise, completely unrestricted calibration may excessively distort the original predictions, adversely affecting the ranking results.

\begin{figure}
    \centering
    \includegraphics[width=0.475\textwidth]{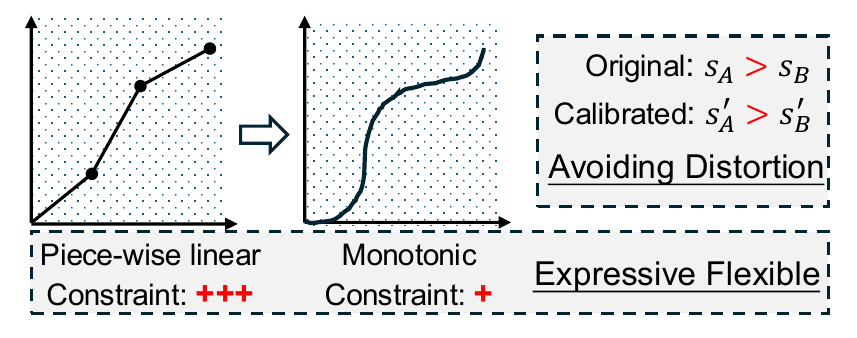}
    \caption{
    The core idea of the proposed monotonic calibrator. 
    It solely imposes a conditional monotonicity constraint on the calibrator, enhancing the expressiveness and flexibility while avoiding excessive distortion of original predictions. Specifically, for samples $A$ and $B$, when conditioning on sample features, if the original scores satisfy $s_A>s_B$, then the calibrated scores will also satisfy $s_A^\prime>s_B^\prime$. }
    \label{fig:diagram}
    \Description{..}
\end{figure}

In this work, we propose imposing only a conditional monotonicity constraint on the calibrator architecture while allowing flexibility in other dimensions. Specifically, we permit any form of transformation function for different samples, with the sole requirement being monotonicity relative to the original predictions when conditioned on sample features. Compared to existing methods, this approach significantly reduces constraints on the calibrator while maintaining the ability to avoid excessively distorting the original predictions through the monotonicity requirement, enabling an appropriate enhancement of expressiveness and flexibility (\textit{c.f.,} Figure~\ref{fig:diagram}). To achieve this, we propose implementing the calibrator with Unconstrained Monotonic Neural Network (UMNN)~\cite{UMNN}. This neural network architecture is capable of learning arbitrary monotonic functions with great modeling power\footnote{UMNN can be roughly considered a specific type of Multi-Layer Perceptron (MLP) network that maintains monotonicity, and notably, arbitrary monotonic functions are limited to those that are differentiable.}~\cite{DGM, NF, NF4PMI}. Additionally, we incorporate sample features into the network learning, making the network feature-specific and thereby allowing samples with different features to have distinct transformation mechanisms.

Taking one step further, we argue that the commonly used calibration optimization techniques may be suboptimal, particularly for our calibrator with greater flexibility. These methods typically leverage the same Binary Cross Entropy Loss (BCELoss)~\cite{NCF} as ranking model optimization, which lacks a specialized design to emphasize the calibration aspect, potentially tending to merely learn a network with properties similar to uncalibrated ranking models~\cite{Yield}. To address this, we propose an additional Smooth Calibration Loss (SCLoss), which partitions samples into groups according to predictions and focuses on minimizing the squared differences between the average values of predictions and labels for each group. This design is inspired by the principle that, for an ideal calibrator, samples with identical predictions should exhibit an equivalence between the predicted values and the posterior probability of the label being positive~\cite{CC4Rec}. SCLoss is crafted to steer the learned network towards conforming to this principle, thereby potentially enhancing the attainment of calibration objectives. Overall, the monotonic calibrator architecture and the calibration-aware learning strategy constitute our framework, referred to as Unconstrained Monotonic Calibration (UMC). We conduct both offline evaluations and online deployments, yielding empirical evidence that attests to its calibration excellence. 

The main contributions of this work are summarized as follows:
\begin{itemize}[leftmargin=*]

    \item We emphasize the necessity for a more expressive and flexible calibrator to address complex calibration scenarios in ranking systems, proposing an implementation grounded in UMNN.

    \item We design a novel loss function to more directly optimize the calibration objectives, enhancing the learning efficacy of the calibrator with greater flexibility.

    \item 
    \sigir{
    We conducted extensive offline experiments, demonstrating that UMC achieves superior calibration performance. Furthermore, A/B test results indicate that UMC leads to significant improvements across multiple business metrics. UMC has been successfully deployed in Kuaishou's video search system\footnote{\url{https://www.kuaishou.com}}, where it now serves the main traffic for hundreds of millions of active users.
    }
    
\end{itemize}
\section{Related Work}
In this section, we examine existing research on the calibration of ranking systems in recommendation and search. The research can be classified into two categories based on the primary focus: calibrator modeling and loss reconciling.

\subsection{Calibrator Modeling}

The primary research focus lies in calibrator modeling, with the objective of developing powerful calibrators to refine original predictions. Early methods predominantly leverage the statistical characteristics of original predictions to perform univariate transformations that ensure global order preservation. Specifically, binning methods~\cite{HB,MBCT} involve partitioning samples into bins and adjusting each sample by assigning a statistical quantity, such as the average predicted probability of its bin, as the calibrated value. Isotonic regression methods~\cite{IR,SIR} optimize squared errors under a non-decreasing constraint to fit a univariate isotonic calibration function. Scaling methods~\cite{PlattScaling,TemperatureScaling,BetaCalib,GammaGauss,ConfCalib} directly fit predefined transformations, like the logistic function, for calibration purposes. In general, the global order preservation of these methods theoretically prevents the calibration process from affecting the ranking performance. However, the constrained parameter space of these transformations limits their effectiveness, particularly in the context of industrial deep ranking models~\cite{LiRank}.

Later works delve into the exploration of utilizing neural networks for calibration, using them to adaptively learn parameters of transformation functions for samples with varying features. For instance, FAC~\cite{NeuralCalib} combines a univariate piece-wise linear model with a field-aware auxiliary neural network. AdaCalib~\cite{AdaCalib} learns isotonic function families to calibrate predictions, guided by posterior statistics. SBCR~\cite{SBCR} introduces a neural piece-wise linear model that integrates sample features directly into the learning of linear weights. However, due to the inadequate fitting performance of piece-wise linear interpolation~\cite{error}, these methods struggle to thoroughly handle multi-field calibration that involves nuanced patterns. DESC~\cite{DESC}, developed concurrently with our approach, replaces piece-wise linear functions with combinations of multiple nonlinear basis functions. Nevertheless, the calibrator expressiveness remains partially constrained, as it lacks guarantees for fitting arbitrary monotonic functions.

\subsection{Loss Reconciling}
Apart from the calibrator modeling aspect, there is also research focused on designing joint optimization strategies to handle point-wise calibration loss and pairwise or list-wise ranking loss, aiming to enhance compatibility between ranking and calibration.

Specifically, CalSoftmax~\cite{ScaleCalib} addresses training divergence issues of the ranking loss and achieves calibrated outputs through the use of virtual candidates. JRC~\cite{JRC} utilizes two logits for click and non-click states to decouple the optimization of ranking and calibration. RCR~\cite{RCR} proposes a regression-compatible ranking approach to balance calibration and ranking accuracy. CLID~\cite{CLID} employs a calibration-compatible list-wise distillation loss to distill the teacher model's ranking ability without destroying the model's calibration ability. 
SBCR~\cite{SBCR} introduces a self-boosted ranking loss that utilizes dumped ranking scores obtained from the online deployed model, facilitating comparisons between samples associated with the same query and allowing for extensive shuffling of sample-level data.
\bym{BBP~\cite{BBP} tackles the issue of insufficient samples for ranking loss by estimating beta distributions for users and items, generating continuously comparable ranking score labels.} However, they primarily focus on reconciling the optimization of ranking and calibration, instead of developing calibrator architectures. 
\section{Task Formulation}\label{section:pre}
In this section, we present a formal formulation of the calibration process within the ranking system.

\vspace{+5pt}
\textbf{Calibration}. Ranking models are designed to learn user-item (or query-document) matching scores for sorting candidates. As ranking models typically emphasize relative orders, they often exhibit discrepancies between the absolute values of predictions and the actual likelihood~\cite{TemperatureScaling,JRC,RCR,Understanding}. 
These discrepancies can negatively impact downstream tasks that necessitate precise absolute value predictions such as advertisement bidding, necessitating the application of calibration techniques. 
Typically, calibration is performed after training the ranking model~\cite{TemperatureScaling}. This process generally involves learning a calibrator model that adjusts the original predictions from the ranking model based on sample features, with the learning typically conducted using a hold-out dataset $\mathcal{D}$~\cite{NeuralCalib,DESC}.

Formally, let $(\bm{x}, y, s)$ denote a sample, where $\bm{x}$ represents the sample features, $y \in \{0,1\}$ is the label, and $s\in[0,1]$ is the predicted matching score from the ranking model, indicating the predicted probability of the sample being positive (\textit{i.e.}, $y=1$). Our target is to use $\mathcal{D}$ to train a calibrator model $g_{\phi}$ that can adjust the original predicted matching score to approach the actual likelihood, parameterized by $\phi$.
Specifically, we seek to achieve
\begin{equation}\label{eq:calibration}
    P(y=1|\bm{x}) \leftarrow s^{\prime} = g_{\phi}(s,\bm{x}; \mathcal{D}),
\end{equation}
where $s^{\prime}\in[0,1]$ is the calibrated prediction and $P(y=1|\bm{x})$ is the actual probability for the sample $(\bm{x},y,s)$.
The absolute values of the calibrated predictions will better align with the actual likelihood, while maintaining or even improving ranking performance, thereby enhancing the overall effectiveness of the ranking system.

\textbf{Ideal Calibration}.
For ideal calibration, each calibrated prediction should match the corresponding actual likelihood, i.e., having $s^{\prime}=P(y=1|\bm{x})$ in Equation~\eqref{eq:calibration}. However, $P(y=1|\bm{x})$ is not directly observable in the real world, so it is challenging to use this criterion to assess the achievement of the ideal state.
Fortunately, a necessary condition for ideal calibration can be inferred from the perspective of posterior probabilities --- samples with identical calibrated predictions $p$ should show an equivalence between the predictions and the posterior probability of label being positive~\cite{CC4Rec}. For instance, if we have 100 samples with $s^{\prime}=0.8$, we would expect that $\frac{80}{100}$ of these samples correspond to $y=1$ if the predicted scores are accurately calibrated. According to~\cite{CC4Rec}, this principle of ideal calibration can be formally expressed as
\begin{equation}\label{eq:ideal-calibration}
\mathbb{E}[y|s^{\prime}=p]=p, \forall{p}\in [0,1],
\end{equation}
where $p \in [0,1]$ represents a possible calibrated score.  
\section{Methodology}

In this section, we first provide an overview of the proposed methodology. Next, we introduce the monotonic calibrator architecture, which offers enhanced calibration expressiveness and flexibility. Subsequently, we describe the proposed calibration-aware learning strategy designed to improve the calibrator's learning efficacy. 

\subsection{Overall Framework}

\begin{figure}[t]
    \centering
    \includegraphics[width=0.495\textwidth]{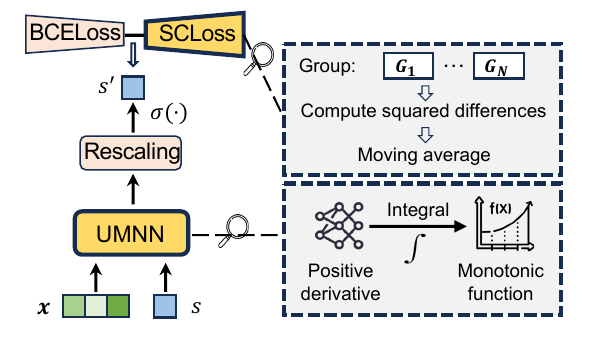}
    \caption{
    \sigir{
    An overview of the proposed UMC framework, which comprises the monotonic calibration architecture (with UMNN module as the core) and the calibration-aware learning strategy (with SCLoss as the core).  The architecture takes the original predictions \( s \) and sample features \( \bm{x} \) as inputs, enforces monotonicity using the positive derivatives of UMNN, and outputs the final calibrated scores \( s^\prime \) through rescaling operations. The learning strategy integrates BCELoss and SCLoss for model optimization, with SCLoss specifically designed to emphasize calibration. 
    }
    }
    \Description{..}
    \label{fig:framework}
    \vspace{-10pt}
\end{figure} 

We strive to enhance the expressiveness and flexibility of the calibrator to improve the calibration effectiveness. Our belief is to relax the constraints on the calibrator by solely constraining conditional monotonicity relative to original predictions while allowing flexibility in other dimensions. To implement it, we propose a novel Unconstrained Monotonic Calibration (UMC) framework, which involves a specifically designed monotonic calibrator architecture and calibration-aware learning strategy as shown in Figure~\ref{fig:framework}.

\begin{itemize}[leftmargin=*]
    \item \textbf{Monotonic Calibrator Architecture}: 
    To satisfy reduced constraint requirements, an ideal calibrator architecture should be capable of learning any monotonic function. With this in mind, the calibrator is built upon an Unconstrained Monotonic Neural Network (UMNN)~\cite{UMNN}, which has similar capabilities to MLPs in fitting any function while ensuring monotonicity~\cite{DGM, NF, NF4PMI}. By feeding the original predictions and sample features into the network, along with some rescaling operations, we enable unconstrained monotonic transformations tailored to the sample features to adjust the original predictions.
     
    \vspace{+5pt}
    \item \textbf{Calibration-aware Learning Strategy}: 
    To effectively train the calibrator, we introduce an additional loss function, referred to as Smooth Calibration Loss (SCLoss), rather than relying solely on the commonly used BCEloss. This new loss function is specifically designed to emphasize fulfilling the necessary condition for achieving the ideal calibration state, as described in Equation~\eqref{eq:ideal-calibration}. We directly transform this equation into a loss format and minimize it to optimize the calibrator, ensuring that it approaches the ideal state from the perspective of necessary conditions.
\end{itemize}

\subsection{Monotonic Calibrator Architecture}

Overall, the monotonic calibrator consists of a UMNN module followed by a rescaling layer, as illustrated on the left side of Figure~\ref{fig:framework}. The UMNN serves as the core module for learning monotonic functions, while the rescaling layer enhances the adaptability. We now elaborate on each of these components.

\vspace{+5pt}
\textbf{UMNN}. 
Directly using a neural network to represent a monotonic function can be challenging. Instead, UMNN leverages neural networks to model partial derivatives related to the function and then performs a definite integral to recover the original monotonic function. The key idea is that \textit{if a function is differentiable, its derivative being single-sign is a necessary and sufficient condition for the function to be monotonic.} To ensure monotonicity, we just need to constrain the output of the network that models the derivatives accordingly. 
Formally, we derive a monotonic model $U(s, \bm{x})$ for a sample with features $\bm{x}$ and original prediction $s$ as
\begin{equation}\label{eq:UMNN}
U(s, \bm{x}) = \int_{t=0}^{t=\sigma^{-1}(s)} h(t,\bm{x}) dt + \beta. 
\end{equation}
Here, $h(t, \bm{x})$ represents a single-sign function modeling the partial derivative, $t$ is a variable associated with $s$, $\beta$ is a scalar offset added to the integral, and $\sigma^{-1}(\cdot)$ denotes the inverse of the Sigmoid function, connecting $s$ and $t$. Notably, the monotonicity of $U(\cdot)$ is conditioned on the sample features, facilitating feature-specific transformations when calibration. 
\bym{Meanwhile, the integral computation can be efficiently executed using the static Clenshaw–Curtis quadrature algorithm\footnote{\new{The efficiency will be discussed in detail in Section~\ref{sec:online}.}}~\cite{CC}, with a predefined number of integration steps not exceeding 100~\cite{UMNN}.}

\textit{Implementation of $h(\cdot)$}. We implement the function $h(\cdot)$ using an MLP module with a constraint to ensure single-sign outputs. Specifically, we restrict its output to be positive, ensuring that the resulting $U(\cdot)$ is monotonically increasing to $s$ conditioned on sample features $\bm{x}$. The consideration is that ensuring a monotonic increase helps better prevent excessive distortion of the original rankings by maintaining the relative order for samples with the same features. Formally, it can be represented as
\begin{equation}
h(t,\bm{x}) = 1 + \text{ELU}(\text{MLP}([t;\bm{x}])),
\end{equation}
where $[t;\bm{x}]$ denotes the concatenation of $t$ and $\bm{x}$, and $\text{ELU}(\cdot)$ denotes the Exponential Linear Unit ~\cite{ELU} that sets its internal parameter $\alpha=1$. Since the output of $\text{ELU}(\cdot)$ is always greater than -1, $h(t,\bm{x})$ will consistently be positive.

\textit{Monotonicity proof.} It is easy to verify that the partial derivative of $U(s,\bm{x})$ to $s$ satisfy that 
\begin{equation}\label{eq:derivative} 
\frac{d}{ds} U(s,\bm{x})=h(\sigma^{-1}(s),\bm{x}) \cdot (\sigma^{-1}(s))' >0,
\end{equation}
where $(\sigma^{-1}(s))'$ denotes the derivative for $\sigma^{-1}(s)$.
This confirms that the calibrated prediction is monotonic relative to the original prediction when conditioned on sample features.

\vspace{+5pt}
\textbf{Rescaling}. 
After obtaining the output $U(s,\bm{x})$ from UMNN, a rescaling operation is conducted to adjust its scale and convert the result to represent probabilities. This operation could enhance the overall network learnability~\cite{ExU}, enabling a more precise capture of intricate mapping patterns.
Specifically, the features $\bm{x}$ are fed into a distinct MLP module to compute a feature-specific rescaling factor $w(\bm{x})$ and bias term $b(\bm{x})$, help adaptively adjust outputs based on features. Then, the Sigmoid function is employed to convert the result to the probabilistic output, obtaining the calibrated result $s^{\prime}\in[0,1]$. Finally, our calibrator can be formulated as
\begin{equation}\label{eq:final}
g_{\phi}(s,\bm{x}) = \sigma(e^{w(\bm{x})}\cdot U(s,\bm{x})+ b(\bm{x})).
\end{equation}
Here, $w(\bm{x})$ undergoes exponentiation to preserve the monotonic nature of UMNN, and $\phi$ denotes all parameters of our calibrator including those from UMNN and rescaling network.

\subsection{Calibration-aware Learning Strategy}

When optimizing the calibrator, in addition to using the commonly employed BCELoss, we introduce a Smooth Calibration Loss (SCLoss). Unlike BCELoss, the SCLoss is specifically designed to enforce the calibrator to meet the ideal calibration principles discussed in Section~\ref{section:pre}, ensuring the achievement of the ideal calibration state. Next, we present the detailed loss design and subsequently summarize the overall learning process.

\vspace{+5pt}
\textbf{SCLoss.} 
A straightforward approach to building our loss is to directly minimize the squared differences between the left-hand side, $\mathbb{E}[y|s^{\prime}=p]$, and the right-hand side, $p$, in Equation~\eqref{eq:ideal-calibration}. However, this can be challenging due to the continuous nature of probability values $p$ within the interval $[0,1]$, contrasted with the finite number of available samples.
To address this, we propose an approximate method based on discretization.
Firstly, we discretize the interval of possible values into $N$ equal-width bins. Then, we group together the samples whose calibrated predictions fall within the same bin, obtaining $N$ groups of samples. Finally, we construct our SCLoss by calculating the squared differences between the average values within each group and performing a weighted average based on the group sizes. Formally, given a batch of data $\mathcal{B}$, the proposed SCLoss can be defined as
\begin{equation}\label{eq:SCLoss}
\begin{split}
     &\text{SCLoss}(\mathcal{B})=\frac{1}{|\mathcal{B}|}\sum_{k=1}^{N} {|{G}_{k}|} (\bar{y}_{k}-\bar{s}^{\prime}_{k})^2,\\
     &{G}_{k} = \{(\bm{x},y,s)\in \mathcal{B} \, | \, \frac{k-1}{N} \le s^{\prime} < \frac{k}{N} \},
\end{split}
\end{equation}
where ${G}_{k}$ is the $k$-th group data, $s^{\prime}\in[0,1]$ is obtained by the calibrator $g_{\phi}(s,\bm{x})$, $\bar{y}_{k}$ and $\bar{s}^{\prime}_{k}$ denote the average of
ground-truth labels and calibrated predictions within group ${G}_{k}$, respectively.

Notably, directly computing $\bar{y}_{k}$ and $\bar{s}^{\prime}_{k}$ may exhibit significant instability due to the relatively small size of each data batch. To enhance stability and approximate the global objective, we use the exponential moving average technique~\cite{MetaBalance} to smooth the computation. Specifically, we smooth the result by keeping a fraction of the value from the previous batch. Formally, given the smoothed average values of the $k$-th group from the last batch and $G_k$ from a new batch, the updated averages are computed as
\begin{equation}\label{eq:EMA}
\begin{split}
    \bar{y}_k &\leftarrow \tau \cdot \bar{y}_k + (1-\tau) \cdot \frac{1}{|G_{k}|} \sum_{(\bm{x},y,s)\in G_{k}} y, \\
    \bar{s}^{\prime}_k &\leftarrow \tau \cdot \bar{s}^{\prime}_k + (1-\tau) \cdot \frac{1}{|G_{k}|}\sum_{(\bm{x},y,s)\in G_{k}} s^{\prime}, 
\end{split}
\end{equation}
where the hyper-parameter $\tau\in [0,1]$ is used to control the decay rate in the moving average. The incorporation of the smoothing enables more thorough sample utilization, thereby enhancing the stability of the training process~\cite{AdaTask,MetaBalance}. 

\vspace{+5pt}
\textbf{Overall Loss.} Finally, to optimize the calibrator, we combine the introduced SCLoss and the BCELoss in a weighted sum, which can be formally expressed as
\begin{equation}\label{eq:Loss}
    \mathcal{L}(\mathcal{B}) = \text{BCELoss}(\mathcal{B}) + \lambda \cdot \text{SCLoss}(\mathcal{B}),
\end{equation}
where $\lambda$ is to balance the above two loss components. Note that the BCELoss is computed using average reduction for calibrated predictions and ground-truth labels. When $\lambda=0$, only BCELoss is active, as per the configurations of other methods. 

\begin{algorithm}[t]
    
    \caption{Calibration-aware Learning Strategy}
    \label{alg:UMC}
    \LinesNumbered
    \KwIn{calibrator model $g_\phi$, hold-out dataset $\mathcal{D}$, group number $N$, decay rate $\tau$, balancing weight $\lambda$}
    Initialize calibrator model parameters $\phi$\;
    \While{stop condition is not reached}{
        Initialize $\bar{y}_k$ and $\bar{s}^{\prime}_{k}$ as zero\;
        \For{$t=1,\dots, T$}{
        // Step 1 (computation of calibration outputs)\;
        Sample batch data $\mathcal{B}$ from $\mathcal{D}$\;
        Compute $s^{\prime}$ with Equation~\eqref{eq:final}\;
        // Step 2 (computation of loss functions)\;
        Compute $\bar{y}_k$ and $\bar{s}^{\prime}_{k}$ with Equation~\eqref{eq:EMA}\;
        Compute SCLoss($\mathcal{B}$) with Equation~\eqref{eq:SCLoss}\;
        Combine $\mathcal{L}(\mathcal{B})$ with Equation\eqref{eq:Loss}\;
        // Step 3 (update of calibration parameters)\;
        Backpropagate $\mathcal{L}(\mathcal{B})$ and update $\phi$\;
        }
    }
    return $g_\phi$    
\end{algorithm}

\vspace{+5pt}
\textbf{Learning Process.} Algorithm~\ref{alg:UMC} summarizes
the detailed learning algorithm of our calibrator. During the learning process, updates are iteratively performed on batches of data. Each iteration begins by computing calibration outputs using both original predictions from the ranking model and the sample features (lines 6-7). Subsequently, the algorithm computes SCLoss and BCELoss on the current batch data, aggregating them into the ultimate loss via a weighted sum (lines 9-11). Finally, calibration parameters are updated based on gradients of the total loss (line 13), which can be completed by optimizers like Adam~\cite{Adam}.
\section{Experiment}
In this section, we conduct a series of experiments to answer the following research questions,

\noindent \textbf{RQ1}: How does UMC perform on real-world datasets compared to existing calibration methods?

\noindent \textbf{RQ2}: What is the impact of the individual components of UMC on its calibration effectiveness?

\noindent \textbf{RQ3}: How do the specific hyper-parameters of UMC influence the ultimate calibration performance?

\noindent \textbf{RQ4}: 
\new{How is the performance and efficiency of UMC when deployed in real-world industry ranking systems?}

\subsection{Experimental Setting}
\subsubsection{Datasets}
\begin{table}[t]
\caption{Statistical details of the evaluation datasets.}
\label{exp:data}
\begin{tabular}{ccccc}
\hline
Dataset & \#Feature & \#Traning & \#Validation & \#Test \\ \hline
Avazu   & 21       & 24.3M    & 8.1M        & 8.1M  \\
AliCCP  & 14       & 51.2M    & 17.1M       & 17.1M \\ \hline
\end{tabular}
\end{table}

We conduct extensive experiments on the following two public datasets,
\begin{itemize}[leftmargin=*]

    \item \textbf{Avazu}. \bym{This is an advertising dataset provided by Avazu corporation in the Kaggle CTR Prediction Contest\footnote{\url{http://www.kaggle.com/c/avazu-ctr-prediction}}, which contains the click data between users and items spanning 10 days and comprises approximately 40 million samples. We use all 21 categorical features during training. }

    \item \textbf{AliCCP}~\cite{ESMM}. This dataset is an e-commerce dataset provided by Alibaba for Click and Conversion Prediction\footnote{\url{https://tianchi.aliyun.com/dataset/408}}, which contains click and conversion data between users and items on Taobao. We utilize the 14 categorical features with the click behavior as the label for training in our experiments.
\end{itemize}

\new{We split the dataset chronologically into training, validation, and test sets in a 3:1:1 ratio. The summary statistics of the preprocessed datasets are presented in Table~\ref{exp:data}. We adopt the common evaluation setting, consistent with previous work~\cite{DESC,AdaCalib,NeuralCalib}. First, we train a base neural network on the training set as the prediction model and fix it. Next, the base model generates predictions for the validation set samples, which are subsequently used to fit (or train) the calibration models. Finally, the test set is utilized to evaluate the performance of the calibration methods.}

\subsubsection{Baselines}
We compare UMC with the following methods. 
\begin{itemize}[leftmargin=*]
    \item \textbf{Uncalib} (Uncalibration). This method involves no calibration and utilizes the original prediction directly.
    \item \textbf{HistBin} (Histogram Binning)~\cite{HB}. This method partitions the samples into bins and adjusts each sample by assigning the average probability of its respective bin as the calibrated value.
    \item \textbf{IsoReg} (Isotonic Regression)~\cite{IR}. This method optimizes squared errors under a non-decreasing constraint to fit a univariate isotonic calibration function.
    \item \textbf{SIR} (Smoothed Isotonic Regression)~\cite{SIR}. This method further introduces linear interpolation strategy based on IR to improve the calibration smoothness.
    \item \textbf{Platt} (Platt Scaling)~\cite{PlattScaling}. This method  directly leverages the logistic function for calibration.
    \item \textbf{Gauss} (Gaussian Scaling)~\cite{GammaGauss}. This method further extends consideration to the distinct distribution of each class based on the Platt Scaling method.
    \item \textbf{FAC} (Field-Aware Calibration)~\cite{NeuralCalib}. This method combines both a simple piece-wise linear model and a field-aware auxiliary neural network for calibration.
    \item \textbf{SBCR} (Self-Boosted Calibrated Ranking)~\cite{SBCR}. This method introduces a neural piece-wise linear model that integrates sample features directly into the learning of linear weights.
    \item \textbf{DESC} (Deep Ensemble Shape Calibration)~\cite{DESC}. This method replaces piece-wise linear functions with multiple nonlinear basis functions and designs an ensemble framework.
\end{itemize}

\new{Notably, the last three deep learning-based calibration methods represent the current state-of-the-art (SOTA) and are also developed using neural network architectures.}

\subsubsection{Evaluation Metrics}
\new{
Our goal is to enhance calibration performance while maintaining ranking quality. \textbf{Thus, in the evaluation process, we prioritize comparing calibration metrics, using ranking metrics as auxiliary indicators for reference.}
}

Ranking metrics include AUC and GAUC~\cite{GradCraft}, respectively assessing global and local ranking performance. 
For calibration assessment, we employ widely-used metrics including \textbf{ECE} (Expected Calibration Error)~\cite{TemperatureScaling}, \textbf{FRCE} (Field Relative Calibration Error)~\cite{NeuralCalib}, and \textbf{MFRCE} (Multi-Field Relative Calibration Error)~\cite{DESC}. Specifically, ECE quantifies the calibration performance using the sum of absolute differences within equal-width bins over the interval $[0, 1]$, which can be expressed as
\begin{equation}\label{eq:ECE}
    \text{ECE} = \frac{1}{|\mathcal{D}|} \sum_{k=1}^{M}|\sum_{(\bm{x},y,s)\in\mathcal{D}} (s^{\prime}-y ) \cdot \mathbb{I}({ s^{\prime} \in [\frac{k-1}{M}, \frac{k}{M})})|,
\end{equation}
where $\mathcal{D}$ is the evaluation dataset, $M$ is the binning number, $s^{\prime}\in[0,1]$ is the calibrated prediction, and $\mathbb{I}(\cdot)$ represents the binary indicator function.  FRCE measures calibration within a specific field $z$, and MFRCE computes the average FRCE across multiple fields, which can be expressed as
\begin{equation}\label{eq:FRCE}
    \text{FRCE} = \frac{1}{|\mathcal{D}|} \sum_{k}\frac{|\sum_{(\bm{x},y,s)\in\mathcal{D}} (s^{\prime}-y ) \cdot \mathbb{I}({z=z_k})|}{|\sum_{(\bm{x},y,s)\in\mathcal{D}} y \cdot \mathbb{I}({z=z_k})|},
\end{equation}
\begin{equation}\label{eq:MFRCE}
    \text{MFRCE} = \frac{1}{N_z} \sum_{z} \text{FRCE},
\end{equation}
where $z_k$ represents a particular value of the field, and $N_z$ is the number of selected fields. In accordance with ~\cite{SBCR}, the binning number $M$ for ECE is set to 100. For FRCE, the distinctive field is designated as "site\_id" for Avazu and "101" for AliCCP. Regarding MFRCE, all categorical fields are evaluated across each dataset.

\begin{table*}[t]
\caption{
\sigir{
Performance comparison between the baselines and our UMC on the Avazu and AliCCP datasets, with the best results in bold and sub-optimal results underlined. Notably, the calibration metrics ECE, FRCE, and MFRCE are the primary evaluation metrics, while the ranking metrics AUC and GAUC are used as reference to assess whether the evaluated method enhances calibration performance without compromising ranking quality. Higher values indicate better AUC and GAUC, while lower values are better for other metrics.
}
}
\label{exp:main}
\scalebox{1.1}{
\begin{tabular}{cccccclccccc}
\hline
                          & \multicolumn{5}{c}{Avazu}                                                                                                                                                      &  & \multicolumn{5}{c}{AliCCP}                                                                                                                                                     \\ \cline{2-6} \cline{8-12} 
\multirow{-2}{*}{Method}  & AUC$\uparrow$      & GAUC$\uparrow$     & \cellcolor[HTML]{EFEFEF}ECE$\downarrow$    & \cellcolor[HTML]{EFEFEF}FRCE$\downarrow$   & \cellcolor[HTML]{EFEFEF}MFRCE$\downarrow$  &  & AUC$\uparrow$      & GAUC$\uparrow$     & \cellcolor[HTML]{EFEFEF}ECE$\downarrow$    & \cellcolor[HTML]{EFEFEF}FRCE$\downarrow$   & \cellcolor[HTML]{EFEFEF}MFRCE$\downarrow$  \\ \hline
Uncalib                   & 0.7413             & 0.6544             & \cellcolor[HTML]{EFEFEF}0.0413             & \cellcolor[HTML]{EFEFEF}0.4207             & \cellcolor[HTML]{EFEFEF}0.3484             &  & 0.6348             & 0.5782             & \cellcolor[HTML]{EFEFEF}0.0202             & \cellcolor[HTML]{EFEFEF}1.4824             & \cellcolor[HTML]{EFEFEF}0.7291             \\
HistBin~\cite{HB}         & 0.7309             & 0.6368             & \cellcolor[HTML]{EFEFEF}0.0090             & \cellcolor[HTML]{EFEFEF}0.2701             & \cellcolor[HTML]{EFEFEF}0.2035             &  & 0.5239             & 0.5068             & \cellcolor[HTML]{EFEFEF}0.0076             & \cellcolor[HTML]{EFEFEF}1.2112             & \cellcolor[HTML]{EFEFEF}0.3909             \\
IsoReg~\cite{IR}          & 0.7413             & 0.6544             & \cellcolor[HTML]{EFEFEF}0.0088             & \cellcolor[HTML]{EFEFEF}0.2313             & \cellcolor[HTML]{EFEFEF}0.1655             &  & 0.6348             & 0.5782             & \cellcolor[HTML]{EFEFEF}0.0078             & \cellcolor[HTML]{EFEFEF}1.0168             & \cellcolor[HTML]{EFEFEF}0.3555             \\
SIR~\cite{SIR}            & 0.7413             & 0.6544             & \cellcolor[HTML]{EFEFEF}0.0083             & \cellcolor[HTML]{EFEFEF}0.2147             & \cellcolor[HTML]{EFEFEF}0.1478             &  & 0.6348             & 0.5782             & \cellcolor[HTML]{EFEFEF}0.0080             & \cellcolor[HTML]{EFEFEF}1.0751             & \cellcolor[HTML]{EFEFEF}0.3721             \\
Platt~\cite{PlattScaling} & 0.7413             & 0.6544             & \cellcolor[HTML]{EFEFEF}0.0109             & \cellcolor[HTML]{EFEFEF}0.2294             & \cellcolor[HTML]{EFEFEF}0.1606             &  & 0.6348             & 0.5782             & \cellcolor[HTML]{EFEFEF}0.0078             & \cellcolor[HTML]{EFEFEF}1.0179             & \cellcolor[HTML]{EFEFEF}0.3560             \\
Gauss~\cite{GammaGauss}   & 0.7413             & 0.6544             & \cellcolor[HTML]{EFEFEF}0.0095             & \cellcolor[HTML]{EFEFEF}0.2346             & \cellcolor[HTML]{EFEFEF}0.1640             &  & 0.6348             & 0.5782             & \cellcolor[HTML]{EFEFEF}0.0078             & \cellcolor[HTML]{EFEFEF}1.0180             & \cellcolor[HTML]{EFEFEF}0.3561             \\
FAC~\cite{NeuralCalib}    & 0.7515             & 0.6649             & \cellcolor[HTML]{EFEFEF}\underline{0.0042} & \cellcolor[HTML]{EFEFEF}0.1473             & \cellcolor[HTML]{EFEFEF}\underline{0.0974} &  & 0.6358             & 0.5791             & \cellcolor[HTML]{EFEFEF}0.0065             & \cellcolor[HTML]{EFEFEF}0.9758             & \cellcolor[HTML]{EFEFEF}0.3186             \\
SBCR~\cite{SBCR}          & \underline{0.7516} & \underline{0.6651} & \cellcolor[HTML]{EFEFEF}0.0051             & \cellcolor[HTML]{EFEFEF}\underline{0.1416} & \cellcolor[HTML]{EFEFEF}0.0994             &  & 0.6362             & 0.5797             & \cellcolor[HTML]{EFEFEF}\underline{0.0063} & \cellcolor[HTML]{EFEFEF}\underline{0.9536} & \cellcolor[HTML]{EFEFEF}\underline{0.3096} \\
DESC~\cite{DESC}          & 0.7514             & 0.6637             & \cellcolor[HTML]{EFEFEF}0.0074             & \cellcolor[HTML]{EFEFEF}0.1549             & \cellcolor[HTML]{EFEFEF}0.1086             &  & \underline{0.6364} & \textbf{0.5802}    & \cellcolor[HTML]{EFEFEF}0.0067             & \cellcolor[HTML]{EFEFEF}0.9697             & \cellcolor[HTML]{EFEFEF}0.3200             \\ \hline
UMC (ours)                & \textbf{0.7521}    & \textbf{0.6654}    & \cellcolor[HTML]{EFEFEF}\textbf{0.0033}    & \cellcolor[HTML]{EFEFEF}\textbf{0.1172}    & \cellcolor[HTML]{EFEFEF}\textbf{0.0837}    &  & \textbf{0.6367}    & \underline{0.5801} & \cellcolor[HTML]{EFEFEF}\textbf{0.0058}    & \cellcolor[HTML]{EFEFEF}\textbf{0.9422}    & \cellcolor[HTML]{EFEFEF}\textbf{0.2951}    \\ \hline
\end{tabular}
}
\end{table*}

\subsubsection {Implementation Details}
To ensure fair comparisons, we employ the DeepFM~\cite{DeepFM} model as the backbone ranking model for all the methods under consideration, following the evaluation approach in previous works \cite{NeuralCalib,DESC}. The hidden layer configuration for the ranking model is set to 512 × 256 × 128, and the embedding size is consistently set to 16 for all features. For our UMC, the calibrator takes all categorical features as inputs, where the hidden layers of the MLP within the UMNN and the rescaling layer are respectively set to 50 × 50 and 200 × 200. This configuration is significantly lighter than the ranking model, aligning more closely with real industrial scenarios and ensuring the comparability of parameter spaces across various calibration methods. 

In terms of calibration optimization, we employ the Adam optimizer~\cite{Adam}, setting the maximum number of epochs to 200. We leverage the grid search to find the best hyper-parameters. 
Considering the scale of both datasets, we set the size of mini-batch to 16384 for all methods, searching the learning rate in the range of $\{$1$e$-4, 5$e$-4, 1$e$-3$\}$, and the $L_2$ regularization coefficient in $\{$0, 1$e$-6, 1$e$-5, 1$e$-4, 1$e$-3$\}$. 
Specifically for our method, \bym{we set the number of integration steps to 50, as specified in the original configuration of UMNN}. We search the hyper-parameter $N$ in Equation~\eqref{eq:SCLoss}, which sets the number of bins for SCLoss, in the range $\{5, 10, 20, 40\}$. The hyper-parameter $\tau$ in Equation~\eqref{eq:EMA}, which regulates the smoothness of the moving average, is searched in $\{0.90, 0.95, 0.99\}$. Additionally, the hyper-parameter $\lambda$ in Equation~\eqref{eq:Loss}, which balances the two losses, is searched in $\{$1$e$-3, 1$e$-2, 1$e$-1, 1, 10$\}$. For the specific hyper-parameters of the baseline methods, we search the ranges as defined in their respective papers.

\subsection{Performance Comparison (RQ1)}

\sigir{We begin by evaluating the overall performance of the compared methods in optimizing calibration while maintaining ranking performance.}
The summarized results
are presented in Table~\ref{exp:main}, yielding the following observations,
\begin{itemize}[leftmargin=*]
    \item 
    \new{UMC demonstrates superior calibration performance compared to all the baseline methods across both datasets.
    For example, on the Avazu dataset, UMC achieves a significant improvement over the best baseline method, with enhancements of +21\% in ECE, +17\% in FRCE, and +14\% in MFRCE. 
    \sigir{These results can be attributed to the proposed expressive and flexible monotonic calibrator, along with its effective calibration-aware learning strategy.}
    By contrast, other neural network-based baseline methods (FAC, SBCR, and DESC) exhibit significantly inferior calibration performance, compared to UMC. Their limitations primarily arise from reliance on piecewise linear fitting paradigms or combinations of nonlinear basis functions, which constrain the expressiveness of their calibrator models.}

    \item 
    \sigir{
    Regarding ranking performance, all neural network-based methods (FAC, SBCR, DESC, and our UMC) generally yield comparable results across both datasets, with our method performing slightly better in most cases. Combined with UMC's superior calibration performance, this further validates the superiority of UMC. Additionally, neural network-based methods outperform traditional methods,  
    which can be attributed to the enhanced model capacity of neural network architectures. 
    }

    \item Uncalib exhibits the poorest performance in both ranking and calibration metrics, highlighting the need for the calibration stage after the training of the ranking model. HistBin exacerbates the ranking metrics after applying calibration, which can be attributed to its indiscriminate assignment of the same calibration output to samples within the same bin, emphasizing the importance of flexible calibration.
    
    \item Other methods, including IsoReg, SIR, Platt and Gauss, strictly maintain ranking metrics and to some extent reduce calibration errors. Besides, they exhibit inferior calibration performance compared to neural network-based methods. This disparity stems from their reliance on univariate mapping functions, which underscores the necessity for enabling distinct transformation mechanisms for samples with varying features.

\end{itemize}

\subsection{Ablation Study (RQ2)}


\begin{table}[t]
\caption{Results of the ablation study for UMC on Avazu.}
\label{exp:abl}
\begin{tabular}{lccccc}
\hline
Method        & AUC$\uparrow$ & GAUC$\uparrow$ & ECE$\downarrow$ & FRCE$\downarrow$ & MFRCE$\downarrow$ \\ \hline
Full UMNN           & 0.7521        & 0.6654         & 0.0033          & 0.1172           & 0.0837            \\
w/o Rescaling & 0.7508        & 0.6630         & 0.0078          & 0.1558           & 0.1118            \\
w/o Feature   & 0.7413        & 0.6544         & 0.0080          & 0.2119           & 0.1465            \\
w/o UMNN      & 0.7413        & 0.6544         & 0.0095          & 0.2482           & 0.1792            \\ \hline
Full SCLoss           & 0.7521        & 0.6654         & 0.0033          & 0.1172           & 0.0837            \\
w/o Average   & 0.7508        & 0.6626         & 0.0084          & 0.1613           & 0.1248            \\
w/o SCLoss    & 0.7519        & 0.6651         & 0.0072          & 0.1672           & 0.1154            \\
w/ MSELoss    & 0.7521        & 0.6649         & 0.0061          & 0.1512           & 0.0982            \\ \hline
\end{tabular}
\end{table}

To enhance the calibration performance in UMC, we propose integrating the monotonic calibrator architecture alongside a calibration-aware learning strategy. 
To substantiate the rationale behind these design decisions, we conduct an exhaustive evaluation by systematically disabling one critical design element at a time to obtain various variants. 
For the monotonic calibrator architecture, we introduce the following three model variants for comparison,
\begin{itemize}[leftmargin=*]
    \item \textbf{UMC w/o Rescaling}. This variant removes the rescaling layer of the architecture and relies solely on the UMNN output.
    \item \textbf{UMC w/o Feature}. This variant eliminates the feature input of the UMNN, resulting in a univariate mapping function.
    \item \textbf{UMC w/o UMNN}. This variant replaces the UMNN with a piece-wise linear model in FAC.
\end{itemize}
While for the calibration-aware learning strategy, we introduce the following two model variants,
\begin{itemize}[leftmargin=*]
    \item \textbf{UMC w/o Average}. This variant sets decay rate $\tau$ to zero, removing the exponential moving average usage of SCLoss.
    \item \textbf{UMC w/o SCLoss}. This variant sets balancing wight $\lambda$ to zero, completely disables the effect of SCLoss. 
    \item \textbf{UMC w/ MSELoss}. \bym{This variant replaces SCLoss with MSELoss, aligning the calibrated output with the binary label.} 
\end{itemize}

Table~\ref{exp:abl} illustrates the comparison results on Avazu, from which we draw the following observations,
\begin{figure}[t]
    \centering
    \includegraphics[width=0.45\textwidth]{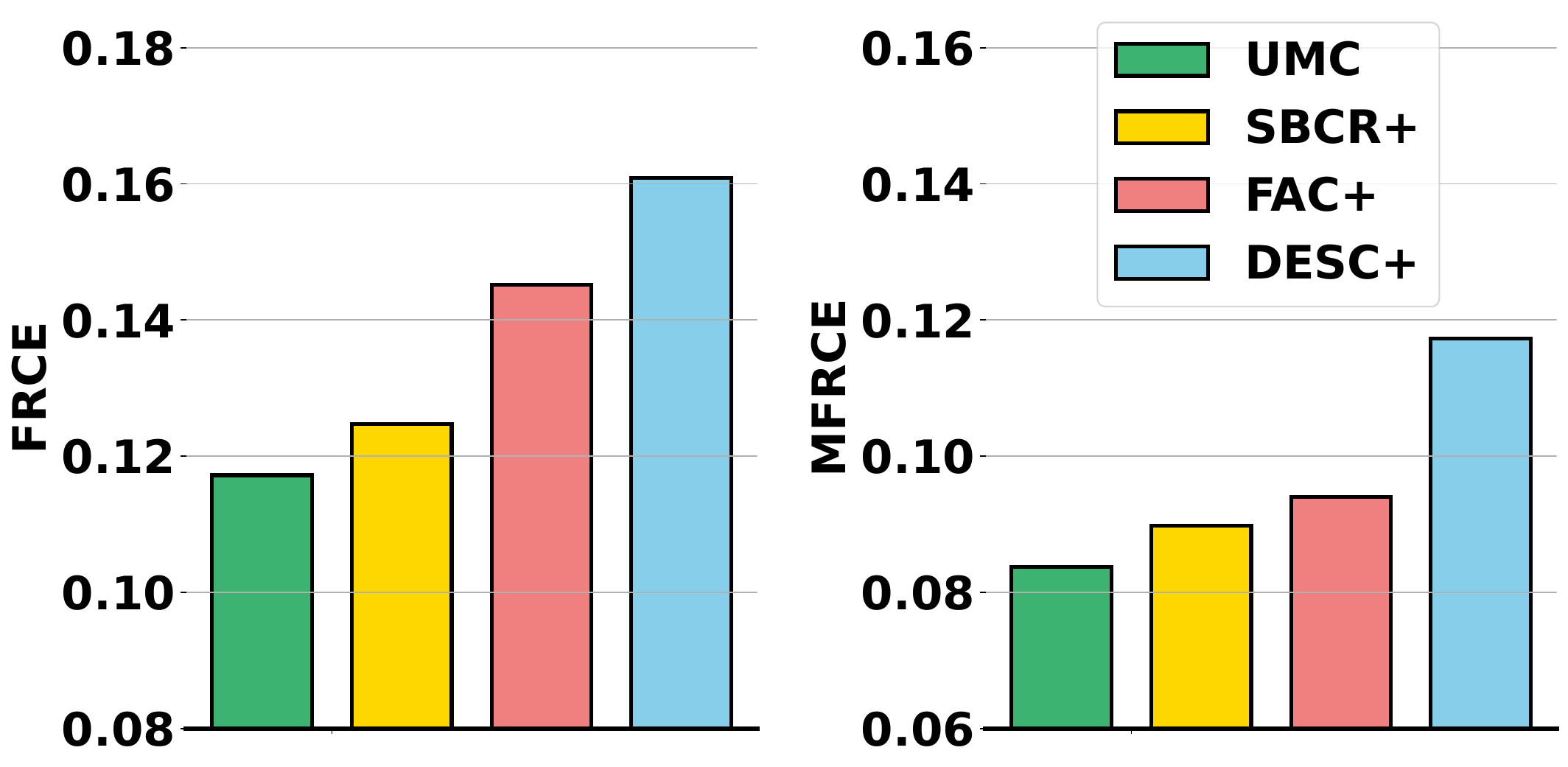}
    \caption{Results of adding SCLoss to other neural network-based baseline methods on Avazu.}
    \Description{..}
    \label{fig:base+}
\end{figure}
\begin{itemize}[leftmargin=*]
    \item The removal of the rescaling layer leads to decreases in all metrics. This discovery underscores the effectiveness of our design in enhancing overall learnability, thereby enabling a more accurate representation of intricate mapping patterns.

    \item Eliminating the feature inputs and substituting the UMNN both result in ranking metrics aligning entirely with Uncalib, as they forfeit feature-awareness capabilities and enforce global order preservation. Besides, replacing UMNN with a piece-wise linear model yields inferior results, validating the expressiveness of our monotonic architecture even in the absence of feature inputs.

    \item Disabling the moving averages usage and discarding SCLoss both result in a performance decline, with the former yielding poorer outcomes. This can be attributed to the computation within each group being restricted to the current batch, which diverges from the global objective and misguides the learning process. 
    These findings underscore the importance of our learning strategy in effectively training the calibrator.

    \item \bym{The introduction of MSELoss provides a performance improvement over solely using BCELoss. However, its effectiveness still falls short of SCLoss, underscoring the unique advantage of SCLoss in enhancing calibration.}
    
\end{itemize}

Building on these findings, we next evaluate the compatibility of our learning strategy with other neural network-based methods. To achieve this, we incorporate our SCLoss into three baseline methods, denoted as \textbf{FAC+}, \textbf{SBCR+}, and \textbf{DESC+}, and tune the weight in Equation~\eqref{eq:Loss}. The comparative results of FRCE and MFRCE are summarized in Figure~\ref{fig:base+}. For the sake of simplicity, the results of ECE, which exhibit a similar trend, are omitted. 
We can observe that while the integration of our learning strategy may enhance the calibration performance of the other methods, a gap remains when compared to UMC. 
This result demonstrates that SCLoss can enhance the achievement of calibration objectives and  unlocking the full potential of our calibrator. 
Furthermore, this finding suggests that our monotonic calibrator is more compatible with SCLoss due to its increased flexibility, thereby offering greater calibration potential for our overall framework.

\subsection{Hyper-parameter Analysis (RQ3)}

\begin{figure}
    \centering
    \includegraphics[width=0.49\textwidth]{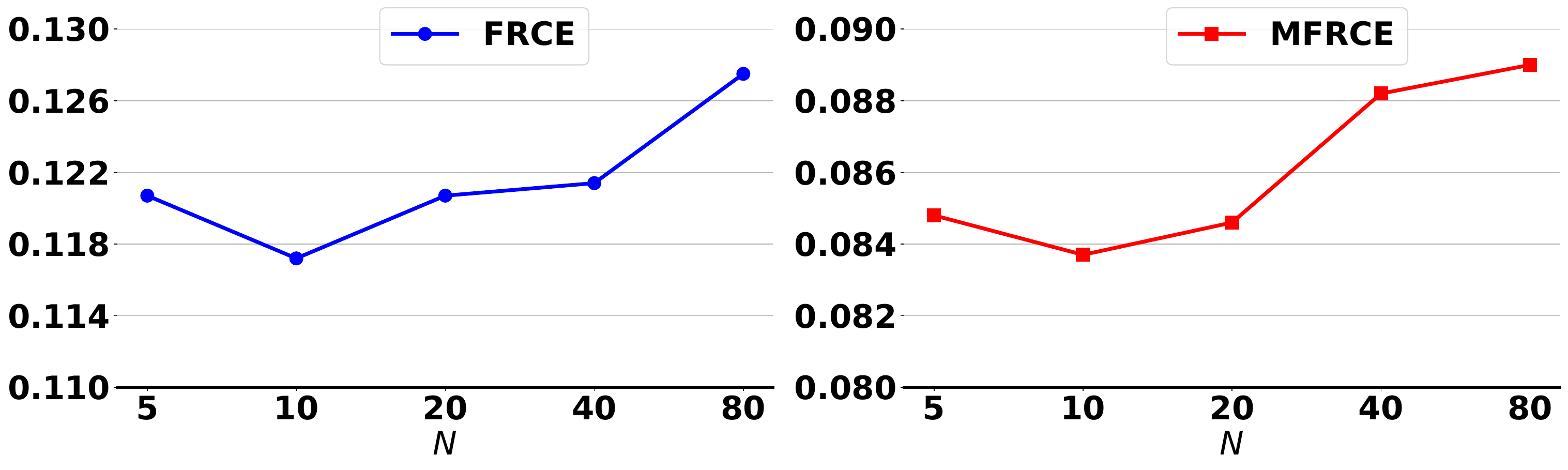}
    \caption{Results of the performance of UMC across different values of group number $N$ on Avazu.}
    \label{fig:hyper_N}
    \Description{..}
\end{figure}

\begin{figure}
    \centering
    \includegraphics[width=0.49\textwidth]{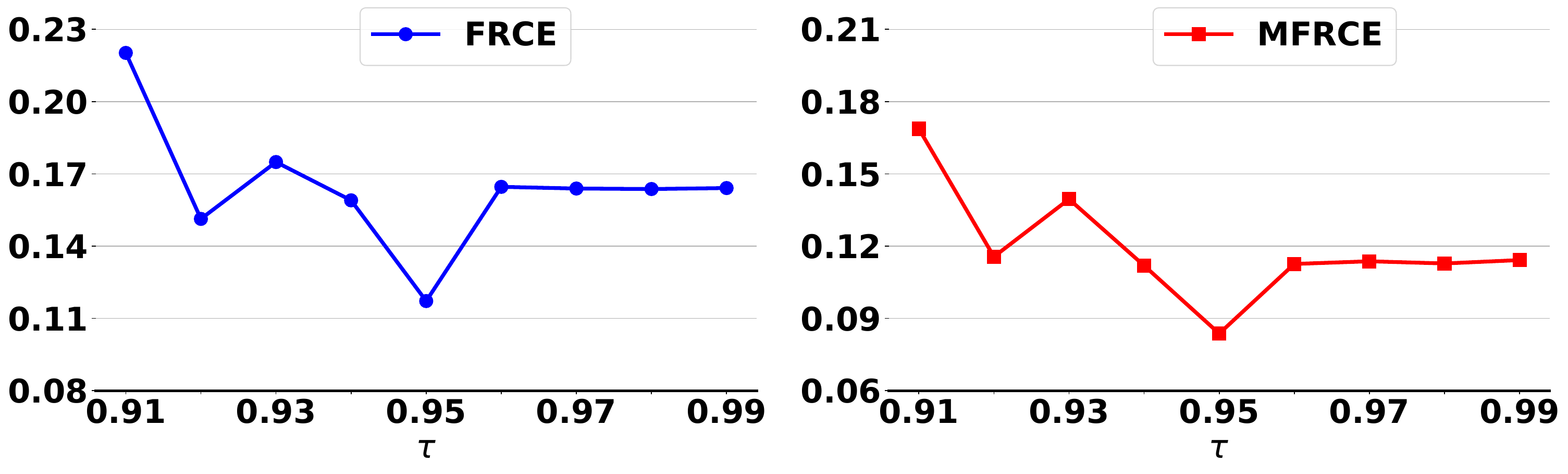}
    \caption{Results of the performance of UMC across different values of decay rate $\tau$ on Avazu.}
    \label{fig:hyper_tau}
    \Description{..}
\end{figure}

In our investigation, the two hyper-parameters $N$ and $\tau$ assume pivotal roles in influencing the
effectiveness of the learning strategy. Specifically, $N$ controls the  granularity of the binning scheme, and $\tau$ governs the extent of smoothing in the moving average computation. We undertake a systematic examination to scrutinize the impact of varying them on the performance. For the group number $N$, we set it in the range of $\{5, 10, 20, 40, 80\}$. For the decay rate $\tau$, we set it from 0.91 to 0.99 with a step size 0.01, as the optimal value is empirically close to one~\cite{MetaBalance}. We summarize the results in Figure~\ref{fig:hyper_N} and Figure~\ref{fig:hyper_tau}, from which we have following observations,

\begin{itemize}[leftmargin=*]
    \item When computing the SCLoss using Equation~\eqref{eq:SCLoss}, a trade-off exists between the granularity and reliability of the binning scheme. Specifically, employing an insufficient number of bins diminishes the ability to differentiate among groups, while an excessive number can lead to bins with sparse data, rendering the computation unreliable. Empirical evidence suggests that selecting 10 bins strikes a balance, ensuring that groups are clearly delineated without compromising the density within each group.

    \item In the computation of the moving average, a value of $\tau$ that is too small may lead to a delayed response to recent batch data, while an excessively large value may obscure long-term trends. Through experimentation, we find that $\tau = 0.95$ achieves an optimal balance between flexibility and stability.
\end{itemize}

\subsection{Online Deployment (RQ4)}\label{sec:online}
\begin{figure}[t]
    \centering
    \includegraphics[width=0.475\textwidth]{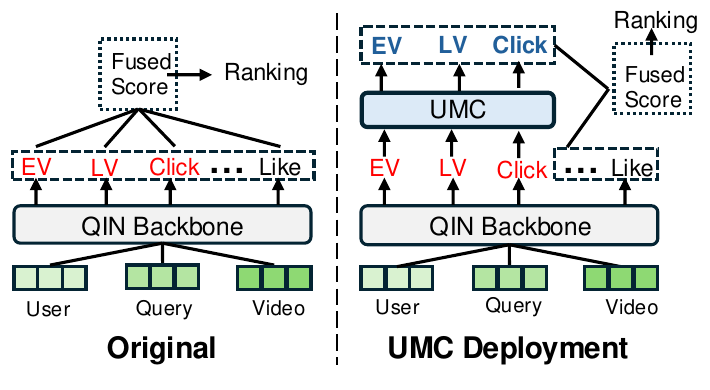}
    \caption{The details of online deployment. UMC calibrates QIN's predictions for EffectiveView (EV), LongView (LV), and Click. The calibrated scores are then used to replace the original scores in the calculation of the fused score, which serves as the final ranking basis.}
    \label{fig:online}
    \vspace{-10pt}
\end{figure}

\new{To further validate the efficacy of UMC, we conduct live A/B experiments on the video search system of Kuaishou -- a representative ranking system that serves over 400 million users daily.}

\vspace{+5pt}
\textbf{Experiment Setup}. The experiment spanned one week, where we randomly split the users into four groups, with proportions of 5\%, 5\%, 5\%, and 85\%. The first three groups are used for online evaluation. Experiments conducted on the 5\% groups impact a significant population of over 20 million users, ensuring the statistical significance of our results. The first group operates without calibration, the second group applies SIR~\cite{SIR} as the calibration method, and the third group employs our proposed UMC for calibration. This comparison setup aligns with previous studies~\cite{DESC, AdaCalib}.
All methods are based on the same backbone, QIN~\cite{QIN}, and share identical features and optimizers. We apply the calibration methods to three key labels in the ranking stage on our platform: EffectiveView~\cite{DML}, LongView~\cite{DML}, and Click. The original predicted scores for these labels are replaced with their calibrated ones, which are then integrated with other scores to rank candidates, as shown in Figure~\ref{fig:online}. For evaluation, we employ widely used metrics in our production systems, including total watch time (WT)~\cite{PEPNet}, effective video views (VV)~\cite{GradCraft}, and click-through rate (CTR)~\cite{SBCR}, which directly capture user satisfaction and retention from a business perspective. 

\vspace{+5pt}
\bym{\textbf{Performance Analysis.}} The results of the online experiment are shown in Table~\ref{exp:online}. 
\new{Compared to the uncalibrated QIN backbone, UMC achieves greater performance improvements than SIR. Specifically, UMC contributes to a +0.414\% increase in WT, a +0.411\% increase in VV, and a +0.170\% increase in CTR compared to SIR. 
\sigir{For these online metrics, an improvement of over 0.1\% is considered significant~\cite{PEPNet}. Notably, with our system serving over 400 million daily users, such improvements can translate into substantial business benefits.}
The improvements can be attributed to the following factors. As illustrated in Figure~\ref{fig:online}, a multi-task learning framework is employed to simultaneously predict multiple labels and integrate their corresponding scores for ranking. 
\sigir{As a result, the calibration of EV, LV, and Click predictions provided by UMC can enhance the quality of the final fused scores.}
This enhancement in both value accuracy and ranking precision ensures better alignment with user interests, ultimately driving higher user engagement.
}

\vspace{+5pt}
\textbf{Efficiency Analysis.} \new{UMC can be efficiently deployed in our system without introducing significant computational overhead.} The primary computational cost may lie in the integral computation within the UMNN component, as described in Equation~\eqref{eq:UMNN}. However, due to the parallelization~\cite{UMNN} of this process, there is no impact on training or inference speed. While memory consumption may slightly increase, it remains within acceptable limits. Specifically, after online deployment, two key efficiency metrics -- QPS (Queries Per Second) and inference time, show only negligible changes: QPS dropped slightly from 4.505k to 4.504k, and inference time increased from 30.43 ms to 30.45 ms. \new{These results validate the scalability of UMC in large-scale industrial systems.} 

\begin{table}[t]
\caption{The improvements of UMC in the online A/B test compared to the baseline calibration method SIR. Here, WT, VV, and CTR stand for total watch time, effective video views, and click-through rate, respectively. It is noteworthy that performance improvements exceeding 0.1\% for these metrics are considered significant. }
\label{exp:online}
\begin{tabular}{cccc}
\hline
Metric          & WT       & VV       & CTR    \\ \hline
Improvement       & \textbf{+0.414\%} & \textbf{+0.411\%} & \textbf{+0.170\%} \\ \hline
\end{tabular}
\end{table}

\section{Conclusion}

This study investigated prediction calibration in industrial ranking systems, proposing an innovative calibration framework. 
From a modeling perspective, we proposed relaxing the constraints on the calibrator architecture and developed a new monotonic calibrator based on UMNN. This architecture significantly enhances calibration flexibility and expressiveness while avoiding excessively distorting the original predictions. From a learning perspective, we introduced SCLoss to guide the calibrator in meeting a necessary condition of ideal calibration, thereby improving the learning efficacy and helping unlock the full potential of our calibrator.

\new{In future work, we plan to extend UMC to other domains and investigate its applicability to various data modalities to evaluate its generalizability. Furthermore, we aim to delve into the theoretical foundations of calibration to elucidate the intricate relationship between ranking and calibration, offering deeper theoretical insights to guide simultaneous improvements in both aspects.}

\begin{acks}
This work is supported by the National Natural Science Foundation of China (62272437) and the CCCD Key Lab of Ministry of Culture and Tourism.
\end{acks}

\bibliographystyle{ACM-Reference-Format}
\balance
\bibliography{8_reference}


\end{document}